\documentstyle[prl,aps]{revtex}
\tighten 
\begin{document}
 
\draft 
\title{The Mu and Tau Number of Supernovae}

\author{C.  J.  Horowitz\footnote{email:  charlie@iucf.indiana.edu} and Gang
Li\footnote{email:  ganli@indiana.edu}} 
\address{Nuclear Theory Center and Department of Physics,\\ 
Indiana University, Bloomington, IN 47405}

\date{\today} 
\maketitle 
\begin{abstract} The neutrino-nucleon cross section is
slightly larger than that for $\bar\nu-N$.  Therefore, $\bar\nu$ will escape
more quickly from core collapse supernovae leaving the star $\nu$ rich.  A
diffusion formalism is used to calculate the time evolution of the mu and tau
lepton number densities.  These quickly reach steady state equilibria.  We
estimate that a protoneutron star with a maximum temperature near 50 MeV 
will contain over 50 \% more $\nu_\mu$ and $\nu_\tau$ than $\bar\nu_\mu$ and
$\bar\nu_\tau$.  Supernovae may be the only known systems with large mu and 
or tau lepton numbers.  
\end{abstract} 
\pacs{97.60.Bw, 11.30.Fs, 95.30.Cq}

Mu and tau neutrinos and antineutrinos are produced copiously in (core 
collapse) supernovae.  These neutrinos with tens of MeV energies can not 
undergo charged current reactions.  Therefore, it is assumed $\mu$ and 
$\tau$ neutrinos and antineutrinos have identical distributions.

However, the cross section for $\nu$-nucleon elastic scattering is somewhat
larger then that for $\bar\nu-N$ scattering.  This allows $\bar\nu$ to escape
the star more easily leaving the supernova $\nu$ rich.  In this paper, we
examine the effects of recoil and or weak magnetism corrections to the $\nu-N$
cross sections.  These corrections are of order $E/M$, where $E$ is 
the neutrino energy and $M$ the nucleon mass.  They lift the degeneracy 
between neutrinos and antineutrinos.  To our knowledge, all previous 
supernovae works (see for example [1]) assume equal $\nu$ and $\bar\nu$ cross 
sections.

We find a large excess of $\nu$ over $\bar\nu$ in the star and a large mu and
tau lepton number for supernovae.  Possible implications of this for the
neutrino signal, sensitivity to new physics, nucleosynthesis and other
phenomena are discussed at the end of this paper.  A diffusion formalism is
used for the neutrino transport.  This is adequate for our purposes since most
of the effect comes from the core of the protoneutron star well inside the
neutrino sphere.

We focus on mu and tau neutrinos.  For these, the physics is very simple and
clear.  We explicitly discuss mu neutrinos.  All of our results apply unchanged
to $\nu_\tau$.  The effect is also present for electron neutrinos.  Indeed, it
will change the ratio of $\nu_e$ to $\bar\nu_e$ and may produce significant
consequences by changing the proton fraction.  However, for $\nu_e$ there are
also charged current reactions so the situation is more complicated.  
Therefore, we postpone a discussion of $\nu_e$ until the end of this paper.

It is a simple matter to expand the $\nu-N$ elastic cross section 
$d\sigma/d\Omega$ to first order in $E/M$, see ref. [2] for example,
$${d\sigma\over d\Omega}={G_F^2E^2\over 4\pi^2}\Bigl\{
\bigl[c_v^2(1+x)+c_a^2(3-x)\bigr]\bigl[1-3{E\over M}(1-x)\bigr]
\pm 4c_a(c_v+F_2){E\over M}(1-x)\Bigr\},\eqno(1a)$$
with $x=$cos$\theta$ for scattering angle $\theta$.
Here the plus sign is for $\nu$ and the minus sign for $\bar\nu$.  The 
vector ($c_v$, $F_2$) and axial ($c_a$) couplings are given in Table I.  
The transport cross section $\sigma=\int d\Omega d\sigma/d\Omega (1-x)$ is,
$$\sigma={2G_F^2E^2\over 3\pi}\Bigl[c_v^2(1-3{E\over M}) +
c_a^2(5-21{E\over M})\pm 8c_a(c_v+F_2){E\over M}\Bigr].  
\eqno(1b)$$ 
The weak anomalous moment $F_2$ describes the $\sigma_{\mu\nu}q^\nu$ 
coupling of the $Z$ to the nucleon.  It could have significant strange quark 
contributions.  Furthermore, the large value of $F_2$ is important for this 
paper.  A measurement of $F_2$ is underway using parity violating electron 
scattering[3].  For simplicity we neglect strange quark contributions to 
$F_2$ and $c_a$ in this paper.  

The difference in transport cross sections, to order $E/M$, is 
$$D={\sigma_\nu - \sigma_{\bar\nu}\over
\sigma_\nu + \sigma_{\bar\nu}} ={8c_a(c_v+F_2)\over 
c_v^2+5c_a^2}{E\over M}=\delta{E\over M}.\eqno(2)$$ 
The coefficient $\delta$ is $3.32$ for neutrons and $2.71$ for protons 
(using Table I).  We assume that nucleon elastic scattering dominates the 
opacity, so $D$ will also give the difference of mean
free paths, $D=(\lambda_{\bar\nu} -
\lambda_\nu)/(\lambda_{\bar\nu} + \lambda_\nu)$.

A simple diffusion equation for neutrinos is,
$${\partial\over\partial t} n(E) -{1\over 3}{\bf \nabla}\cdot \lambda_\nu {\bf
\nabla} n(E)=0.\eqno(3)$$ 
We subtract a similar equation for $\bar\nu$ to get,
$${\partial\over\partial t} [n(E)-\bar n(E)]-{1\over 3} {\bf \nabla}\cdot
\lambda {\bf \nabla} [n(E)-\bar n(E)]= -{1\over 3}{\bf \nabla}\cdot
(\Delta\lambda) {\bf \nabla} [n(E)+\bar n(E)], \eqno(4)$$ 
with $\lambda=(\lambda_{\bar\nu}+\lambda_\nu)/2=\lambda_0 
E_0^2/E^2$ and
$$\Delta\lambda=(\lambda_{\bar\nu}-\lambda_\nu)/2=\lambda D.\eqno(5)$$  
Here the reference mean free path $\lambda_0$ is, 
$$\lambda_0^{-1}={2G_F^2E_0^2\over 3
\pi}(c_v^2+5c_a^2)\rho_n,\eqno(6)$$ 
with $E_0$ an arbitrary reference energy and $\rho_n$ the density of 
neutrons.  For simplicity, $E/M$ terms are dropped in Eq. (6).
Note that we assume pure neutron matter.  A nonzero proton fraction should 
not change our results very much since $\delta$ in Eq. (2) is similar for 
protons and neutrons.

We integrate Eq.(4) over energy $\int d^3E/(2\pi)^3 n(E)=\rho$, assume local
thermodynamic equilibrium and work to lowest order in the $\nu_\mu$ chemical
potential $\mu$ over the temperature $T$, 
$$\rho-\bar\rho={\mu T^2\over 6},\ \ \
\ \ \rho+\bar\rho={ 3\zeta(3)\over 2\pi^2} T^3,\eqno(7a)$$ 
to get, 
$$-{\pi^2\over E_0^2}{\partial\over\partial t}(T^2\mu)+{\bf \nabla}\cdot 
\lambda_0 {\bf \nabla} \mu ={\pi^2\over 6}{\delta\over M} {\bf \nabla}
\cdot\lambda_0 {\bf \nabla} T^2. \eqno(7b)$$ 
This equation describes the time evolution of the muon number density
in the star.  We also calculate the lepton number current, 
$${\bf J}_\nu - {\bf J}_{\bar\nu} = -{\lambda_0E_0^2\over 6\pi^2} 
{\bf \nabla} (\mu -{\pi^2\over 6}{\delta\over M}T^2).\eqno(8)$$ 
The first term in Eq. (8) describes 
the conventional diffusion of lepton number while the second term comes 
from the diffusion of neutrino pairs which produce a lepton number current 
because of the difference in mean free paths $\Delta\lambda$.  In steady 
state equilibrium ${\bf J}_\nu- {\bf J}_{\bar\nu}=0$ giving for the chemical potential 
$$\mu={\pi^2\over 6}{\delta\over M}T^2,\eqno(9)$$ 
a remarkably simple result.

We now discuss numerical solutions of Eq. (7b) to bound the diffusion time
for muon number in a supernova.  To our knowledge there have been no 
previous estimates of this time.  We find that muon number diffuses faster 
than the thermal energy.  This may be because the heat capacity is 
proportional to the large baryon number.  Furthermore, a low energy 
$\nu$ with a long mean free path can effectively transport muon number [4].  
Since diffusion is fast, we will assume steady state equilibrium and use 
Eq. (9) for some later results.

First, consider a uniform star of density $\rho_0=5\times 10^{14}$ g/cm$^3$ 
and (baryon) mass $M_{tot}=1.5$ Solar masses.  This has a radius $R$ of 11.25 
km.  A lower density surface will only speed up diffusion.  Likewise we 
neglect nucleon Pauli blocking corrections to $\lambda_0$ in Eq. (6).  These 
will also increase the mean free path.  Perhaps the most extreme case is for 
the center of the star to be hot.  Muon number must diffuse all the way to 
r=0.  Therefore we consider a simple temperature distribution characteristic 
of the later stages of the protoneutron star cooling (perhaps 10 sec. 
after collapse [1]), 
$$T(r)= T_0(1-M(r)),\eqno(10)$$ 
with $M(r)$ the enclosed mass divided by $M_{tot}$.  For
a uniform density $M=(r/R)^3$.  We neglect the small time dependence of the
temperature during the short simulation.

We start from the initial condition $\mu=0$ inside the star and adopt a 
somewhat arbitrary surface boundary condition that $\mu$ is in equilibrium, 
given by Eq. (9), at the surface for all times.  Our results are not very 
sensitive to this choice.  Figure 1 shows $\mu$ as a function of time for a 
central temperature $T_0$ of 35 MeV.  Muon number diffuses into the center 
of the star so that eventually $\mu(r=0)$ rises to its equilibrium value 
given by Eq. (9).  This rise happens quickly with $\mu$ reaching half of 
its equilibrium value by $t\approx 0.25$ seconds\footnote{Steady state 
equilibrium is reached even faster for a surface peaked temperature 
distribution.}.  In comparison it takes several seconds for the temperature 
distribution to change from surface peaked to one with a maximum at $r=0$ [5].  
We conclude that lepton number diffusion is fast and expect $\mu$ to be near 
its equilibrium value, Eq. (9).

Equation (9) implies a density asymmetry, 
$${\rho-\bar\rho\over \rho +
\bar\rho}={\pi^2\over 9\zeta(3)}{\mu\over T} ={\pi^4\over
54\zeta(3)}{\delta\over M}T=4.98 {T\over M}.\eqno(11)$$ 
This is large, 0.186 at $T=35$ MeV.  Equations (9) and (11) can be understood 
as follows.  The density of $\nu$ rises above that for $\bar\nu$ until the 
larger $\nu$ density times a shorter mean free path balances 
$\bar\rho\lambda_{\bar\nu}$ for antineutrinos. 

We integrate $\rho-\bar\rho$ over the star assuming Eq. (9) and the
temperature profile in Eq. (10).  This gives, 
$$N-\bar N=\int d^3r(\rho-\bar\rho)= {\pi^3\over 135}
{\delta T_0\over M}\Bigl( {RT_0\over \hbar c}\Bigr)^3 f_1.\eqno(12)$$ 
Here $T_0$ is a characteristic maximum temperature
and the profile factor $f_1=15\int_0^Rr^2dr/R^3[T(r)/T_0]^4$ is one for 
Eq. (10) and is expected to be of order unity for other temperature 
profiles.  We also calculate the total number of muon neutrinos, 
$$N+\bar N = \int d^3r (\rho+\bar\rho)={\zeta(3)\over 2\pi} 
\Bigl({RT_0\over \hbar c}\Bigr)^3 f_2,\eqno(13)$$ 
with the profile factor $f_2=12\int_0^R r^2dr/R^3 [T(r)/T_0]^3$.  For
the temperature distribution in Eq. (10), $f_2=1$.

For example at $T_0=50$ MeV we have, 
$$N-\bar N = 9.40 \times 10^{53},\eqno(14)$$
$$N+\bar N = 4.43 \times 10^{54},\eqno(15)$$ 
or $N=2.69\times 10^{54}$ and $\bar N=1.74 \times 10^{54}$.  
This is a remarkably large asymmetry.  The star contains
54 \% more $\nu_\mu$ than $\bar\nu_\mu$.  One might expect $E/M$ 
correction terms to be small.  However, the coefficient $\delta$ is 
large and the temperature, 50 MeV, is high.  We note that the 
asymmetry depends only linearly on $T_0$, see Eq. (11).  Thus a decrease 
in $T_0$ will not decrease the asymmetry greatly.

Equation (14) is the muon lepton number of the supernova.  Furthermore, it also
gives the $\tau$ number.  We predict that supernovae are the only known 
systems with large $\mu$ and or $\tau$ number.  This could impact 
new physics.  For example, matter enhanced $\nu$ oscillations could depend 
on the $\nu$ density [6].  A weak long range force that couples only to 
$\mu$ or $\tau$ number is difficult to observe.  One may be able to use 
supernovae to set limits on such forces.

We have found a large asymmetry for neutrinos recoiling against heavy 
nucleons.  It is possible that the protoneutron star undergoes a transition 
from hadronic to a pure or mixed quark matter phase which could also contain 
strange quarks [7].  In a quark phase, one expects recoil corrections not of 
order $E/M$ but $E/m_q$ where $m_q\approx M/3$ is a constituent quark 
mass.  This  might produce even larger separations of $\nu$ from $\bar\nu$ 
which could lead to an observable signature of quark matter.

Perhaps the simplest model of quark matter is to assume the neutrinos scatter
from nearly free quarks and that the structure of the quarks are simple without
anomalous moments.  Evaluating Eq. (2) for a down or strange quark using the 
couplings in table I gives, $D\approx 1.01E/(M/3)=3.03E/M$.  This is 
comparable to the neutron value $D=3.32E/M$.  The absence of the factor
$F_2$ reduces the quark value and approximately 
canceles the enhancement from the small mass.  Therefore, we do not expect a 
strong sensitivity to a transition to quark matter.  However, this discussion 
does emphasize that the separation between neutrinos and antineutrinos is 
occurring deep in the protoneutron star and may be sensitive to the 
properties of dense matter.

We now discuss many-body corrections to Eq. (2) assuming a hadronic phase.  
The effects on the nucleons of relativistic kinematics, Fermi motion, 
Pauli blocking and nuclear mean fields[8,9] have been studied in ref. [10].
One might think that $M$ in Eq. (2) will be replaced by the Dirac mass 
$M^*<M$ increasing the asymmetry.  However, the anomalous moment term 
$F_2$ is defined with the free nucleon mass, see Eq. (28) of ref. [11].  
Therefore its contributions do not increase.  We find that $D$ in Eq. (2) is 
almost unchanged in the medium.

Burrows and Sawyer [12] argue that RPA correlations will greatly reduce 
cross sections in dense matter.  However, fully relativistic RPA 
calculations using an interaction that is consistent with the equation 
of state give smaller corrections [10,13].  Furthermore, RPA corrections 
should effect $\nu$ and $\bar\nu$ in about the same way.  Therefore, we do 
not expect significant RPA changes to the cross section difference of Eq. (2).

We now discuss the neutrino signal from a supernova.  We divide the time into
three periods.  For a short initial time interval, the star is not in steady
state equilibrium.  During this period, which we estimate may only last for 
tens to hundreds of msec, the star radiates significantly more $\bar\nu$ than 
$\nu$.  Next, the star will radiate equal numbers of $\nu$ and $\bar\nu$ 
in steady state equilibrium for a relatively long period.  Finally, as the 
star cools it must radiate away its lepton number.  Therefore, there will be 
an ending period where more $\nu$ are radiated than $\bar\nu$.  However, the 
fractional difference between $\nu$ and $\bar\nu$ during this time is 
expected to be small because the star cools slowly.

The number of neutrinos in the star at any one time, Eq. (15) is much less 
than the total number of muon neutrinos radiated (of order $10^{57}$).  
Therefore, the first phase, with its excess $\bar\nu$, can only last for a 
fraction of the total time.  One should consider the possibility of 
observing the excess $\bar\nu$ during the initial phase.  However, it may be 
difficult.  Future work should calculate the neutrino signal in more detail.

If the star undergoes a prompt collapse to a black hole, it will take the large
lepton number of Eq. (14) with it.  Assuming lepton number does not couple 
to a long range force, it will simply be lost down the black hole.  However, 
{\it the brief neutrino signal which precedes a prompt collapse could be 
significantly antineutrino rich.}  Depending on the time of the collapse, 
neutrinos from the more symmetric second and third periods may never escape 
the star.

Finally, we consider electron neutrinos.  The neutral current effect that we
have calculated for $\nu_\mu$ and $\nu_\tau$ will also apply to $\nu_e$.  In
addition, charged current reactions have similar weak magnetism corrections.
These terms will increase the $\nu_e$ cross section with respect to 
$\bar\nu_e$.  Thus, the number of $\nu_e$ compared to $\bar\nu_e$ will 
increase in the star.  This could have a significant effect on the 
dynamics of the explosion and or nucleosynthesis by increasing the proton 
fraction.  We will discuss this further in a later work [5].

The separation between neutrinos and antineutrinos involves parity violation.  
If the weak interactions conserved parity, the $\nu$ and $\bar\nu$ cross 
sections would be equal (to lowest order in $G_F$).  Thus, the large mu and 
tau lepton numbers of a supernova can be considered a macroscopic 
manifestation of parity violation.

In this paper we have calculated the effects of recoil or weak magnetism
corrections to $\nu-N$ cross sections.  These are of order the neutrino
energy $E$ over the nucleon mass $M$, $E/M$ and make
the $\nu-N$ cross section larger than that for $\bar\nu-N$.  A diffusion
formalism was used to follow the time evolution of lepton number.  It quickly
reaches a steady state equilibrium where a larger $\nu$ density compensates for
the longer $\bar\nu$ mean free path.  For a maximum temperature near 50 MeV, 
we estimate that protoneutron stars contain over 50 \% more $\nu_\mu$ and
$\nu_\tau$ than $\bar\nu_\mu$ and $\bar\nu_\tau$.  Core collapse supernovae 
may be the only known systems with large mu and or tau lepton numbers.

\bigskip
\bigskip
\begin{table}
  \caption{Vector and axial couplings for neutral current 
     scattering from neutrons, protons and up and down quarks. 
     We assume sin$^2\theta$=0.23 and $g_a=1.26$.}
   \begin{tabular}{ccccc}
    Coupling & $\nu-n$    &  $\nu-p$   & 
            $\nu-u$ & $\nu-d$ \\
     \tableline
    $c_v$         & $-{1\over 2}$  & ${1\over 2}-2{\rm sin}^2\theta$   & 
    ${1\over 2}-{4\over 3}{\rm sin}^2\theta$  & 
    $-{1\over 2}+{2\over 3}{\rm sin}^2\theta$ \\
    $c_a$         & $-{g_a\over 2}$ & ${g_a\over 2}$  &
    ${1\over 2}$  & $-{1\over 2}$ \\
    $F_2$         & -0.972  & 1.029 &  0 & 0
   \end{tabular}
  \label{tableone}
 \end{table}
\bigskip
\bigskip

\vbox to 5.in{\vss\hbox to 8in{\hss 
{\includegraphics{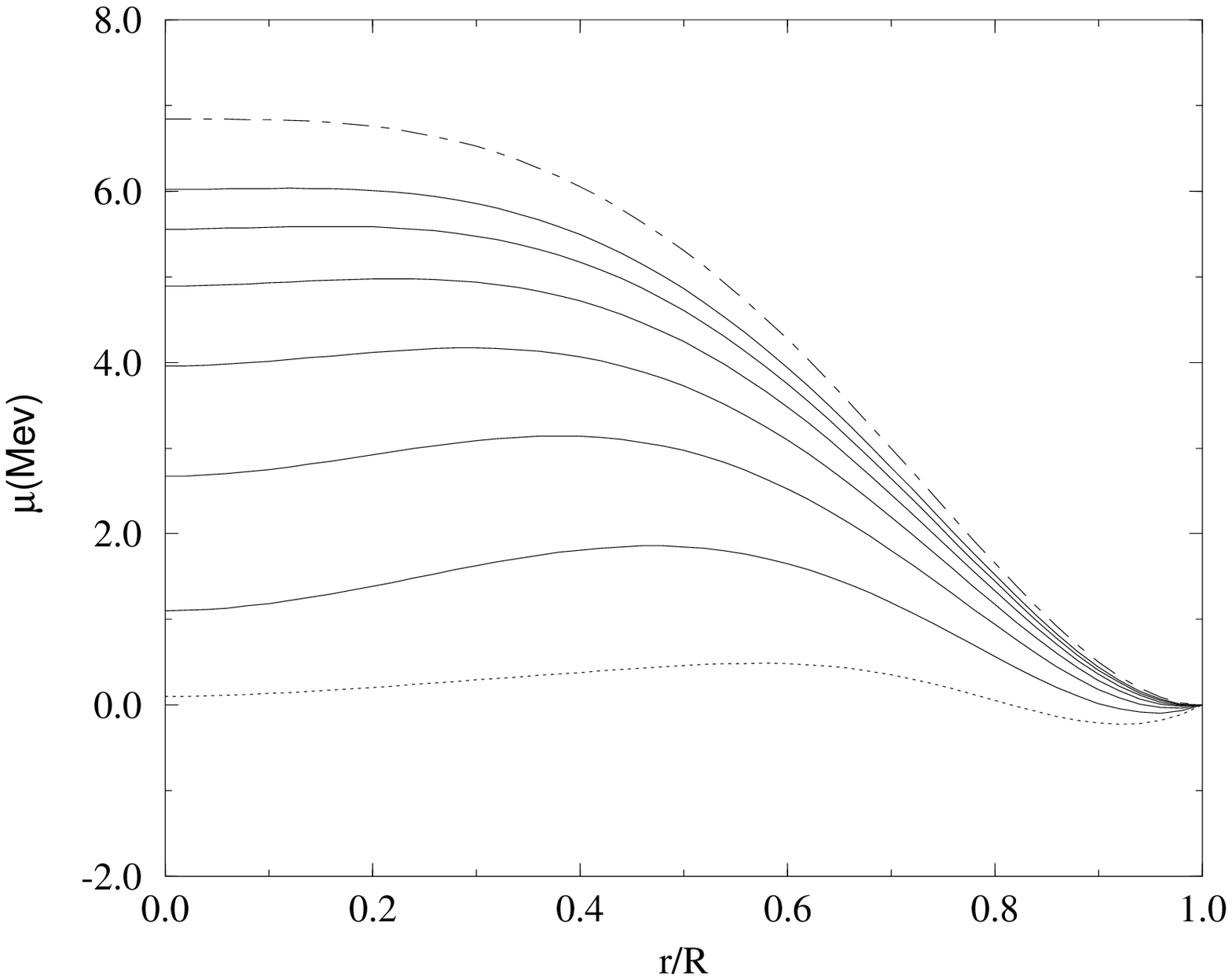}}\hss
}} 
\nobreak
{\noindent\narrower{{\bf FIG.~1}.  Muon neutrino chemical potential vs. 
radius.
The temperature distribution is assumed to be given by Eq. (10) with a
maximum central temperature of $T_0=35$ MeV.  The curves are for times of 
0.02 (dotted) to 1 second (dot-dashed).  The solid curves are for
intermediate times of (bottom to top) 0.1, 0.2, 0.3, 0.4, 0.5, and 0.6 
seconds.
}}

\end{document}